\documentstyle[prl,aps,epsfig,floats]{revtex}
\begin{document}
\draft

\def\lsi{\raise0.3ex\hbox{$<$\kern-0.75em\raise-1.1ex\hbox{$\sim$}}}
\def\gsi{\raise0.3ex\hbox{$>$\kern-0.75em\raise-1.1ex\hbox{$\sim$}}}
\newcommand{\lsim}{\mathop{\lsi}}
\newcommand{\gsim}{\mathop{\gsi}}
\def\tphi{\tilde{\phi}}
\def\tE{\tilde{E}}
\textheight=23.8cm

\twocolumn[\hsize\textwidth\columnwidth\hsize\csname
@twocolumnfalse\endcsname

\title{Phase transitions from preheating in gauge theories}
\author{A. Rajantie
 and E.J. Copeland
}
\address{Centre for Theoretical Physics,
University of Sussex, Falmer, Brighton BN1 9QJ,~~~U.~K. }

\date{\today}
\maketitle
\begin{abstract}
We show by studying the Abelian Higgs model with numerical lattice 
simulations that non-thermal phase transitions arising out of 
preheating after inflation are possible in gauge-Higgs models
under rather general circumstances.
This may lead to the formation of gauged topological defects and,
if the scale at which inflation ends is low enough, to 
electroweak baryogenesis after preheating.
\end{abstract}
\pacs{PACS numbers: 11.15.Ha, 
11.15.Kc, 98.80.Cq \hfill SUSX-TH-00-004
\hfill hep-ph/0003025
}
\vspace*{19.4cm}
\textcircled{c} The American Physical Society
\vspace*{-19.8cm}
]
%\vskip2pc]
%%%%%%%%%%%%%%%%%%%%%%%%%%%%%%%%%%%%%%%%%%%%%%%%%%%%%%

%\section{Introduction}
Over the past few years there has been somewhat of a revolution in our 
understanding of the dynamics of the end of inflation. The traditional 
picture of reheating arising out of the perturbative decay of the inflaton 
field as it oscillates about the minima of its potential has been replaced by 
the possibility of an explosive particle 
production during an earlier period, known as preheating. During preheating, 
parametric 
resonance of the inflaton field generates very large fluctuations of 
the scalar fields coupled to the inflaton, leading to the production of 
large numbers of particles \cite{kofman94}. 
Due to the weakness of the interactions, the  short-wavelength modes
do not thermalize, and the effective temperature of the long-wavelength
modes is much higher than
in the standard reheating scenario. 
This may lead to symmetry restoration and, when the Universe cools
down as it expands further, a subsequent
non-thermal phase transition \cite{kofman96}.   
The fact that the fluctuations produced during preheating
have large occupation numbers implies 
that they 
can be considered as
interacting classical waves, an important result because it means that 
the dynamics of fluctuations during and after preheating can be studied using 
lattice
simulations \cite{khlebnikov96}. 
A concrete example of a non-thermal 
phase transition occuring after preheating 
was presented in Ref.~\cite{khlebnikov98} 
(see also Ref.~\cite{Boyanovsky:1997cr}). 
The phase transition
that they found is first-order, and depending on the field content of the 
model being investigated 
topological defects may form, an intriguing result as it 
opens up the possibility that inflation can create a defect problem
if for example they produce gauged monopoles or domain 
walls~\cite{Kasuya:1997ha,tkachev98,Parry:1999de}.

Non-thermal phase transitions may even solve the old puzzle of baryon 
asymmetry in the Universe~\cite{Rubakov:1996vz}. Although the baryon
number is conserved perturbatively in the Standard Model, there are
non-perturbative interactions that violate this conservation law.
The rate of baryon number violation is extremely low at low energies,
but it becomes much higher in the high-temperature phase of the electroweak 
theory.
Thus it is possible to generate the observed baryon asymmetry
if, for some reason, the fields are out of equilibrium at the electroweak 
scale and thermalize to a temperature
below $T_c$. 
This could be the case even in the standard big bang cosmology,
if the electroweak phase transition were strongly first order, but 
at least in the minimal Standard Model it is not, as lattice simulations 
have shown~\cite{Kajantie:1997qd}. 
However, in a non-thermal phase transition, the fields are driven 
out of equilibrium by the oscillations of the inflaton, and
baryogenesis may
be possible, if the reheating temperature is much lower than 
$T_c$~\cite{Garcia-Bellido:1999sv,Krauss:1999ng}.

Despite the exciting possibility of electroweak baryogenesis, most of the 
numerical
work on non-thermal phase transitions so far has concentrated on scalar
fields \cite{khlebnikov98,Kasuya:1997ha,tkachev98,Parry:1999de} 
or has been restricted to one spatial 
dimension~\cite{Garcia-Bellido:1999sv}.
In this letter we present results from simulations of the Abelian Higgs model
in two rather different cases. The first case is a direct
analogue of the simulations in Ref.~\cite{khlebnikov98}. The inflaton
itself is charged under a gauge group and eventually breaks the gauge
symmetry in a first-order phase transition.
(For simplicity, we use the terminology of spontaneous symmetry breakdown,
although the gauge symmetry is not actually broken in the Higgs
phase.) 
The second case is more relevant for electroweak
baryogenesis. We show that even with a Higgs mass that is compatible
with experimental bounds, the transition is sharp, and electroweak
baryogenesis is therefore possible.
Although we restrict ourselves to the Abelian case in our simulations, 
we expect that 
our conclusions apply to non-Abelian theories as well.

The Lagrangian of our model is
\begin{equation}
L=
-\frac{1}{4}F^{\mu\nu}F_{\mu\nu}
+(D^\mu\phi)^*D_\mu\phi-\lambda(|\phi|^2-v^2)^2.
\label{ownlag}
\end{equation}
Here the gauge covariant derivative is 
$D_\mu\phi=\partial_\mu\phi+ieA_\mu\phi$, and 
$F_{\mu\nu}=A_{\nu;\mu}-A_{\mu;\nu}$.
The couplings $\lambda$ and $e$ are assumed to be small, and
we will use
$\lambda\sim e^2$ in our estimates.

Ideally, we would like to study the quantum field theory defined by 
Eq.~(\ref{ownlag}), but solving for the time evolution of even a simple
quantum system is a formidable task. Therefore we have to resort to
the classical approximation, which is expected to work as long as
the dynamics is determined by modes with a macroscopic occupation
number \cite{khlebnikov96}. 
For studying the dynamics, it is convenient to
fix the temporal gauge $A_0=0$ and 
use the conformal time $\eta$ defined by $d\eta\equiv dt/a$
and the rescaled fields
$\tphi\equiv a\phi$, $\tE_i\equiv-\partial_\eta A_i$.
The equations of motion for $\tphi$ and $A_i$ 
follow from Eq.~(\ref{ownlag}):
\begin{eqnarray}
\partial^2_\eta\tphi &=& D_iD_i\tphi
+(2\lambda v^2a^2+\partial^2_\eta a/a)\tphi
-2\lambda|\tphi|^2\tphi,
\label{omaeqmphi} \\
\partial_\eta \tE_i&=&\partial_jF_{ij}+2e{\rm Im}\tphi^*D_i\tphi,
\label{omaeqmgauge}\\
\partial_i \tE_i&=&2e{\rm Im}\tphi^*\partial_\eta\tphi.
\label{gauss}
\end{eqnarray}

The initial conditions for the fields are those 
produced by inflation: the gauge field is in vacuum and the covariant
derivatives of the Higgs field vanish. This allows us to fix the remaining
gauge degree freedom by setting initially $A_i=0$ and 
$\phi=\overline{\phi}_0=$constant.
%Note that this configuration is a natural consequence
%of inflation, even if $\phi$ itself is not the inflaton. Because of the
%conformal invariance of the gauge sector, the expansion of the
%Universe does not affect the behaviour of the gauge field, and therefore
%the energy in the gauge sector will simply be scaled by the factor 
%$\exp(-Ht)$, i.e.~inflation will dilute all the energy away. 
%The same is true for the inhomogeneous modes of the Higgs field, since
%the timescale of their dynamics is determined by the momentum $p>H$.
%However, the behaviour of the modes with a longer wavelength is different.
%As the Universe expands and the momentum of the mode becomes less than
%$H$, the mode freezes and contributes
%to the homogeneous mode of the final configuration. This effect produces
%for $\phi$ a spatial average of $\overline{\phi}_0\sim(H^3t)^{1/2}$
%or $\overline{\phi}_0\sim(H^2/\lambda)^{1/2}$, whichever is smaller.
%{\bf ?????}

We separate $\phi$ into the homogeneous zero mode $\overline{\phi}$
and the inhomogeneous fluctuations $\delta\phi=\phi-\overline{\phi}$.
We take the quantum nature of the system into account by introducing
small fluctuations
for the fields $A_i$ and $\delta\phi$ and for their canonical
momenta $E_i$ and $\delta\pi\equiv\partial_\eta\delta\phi$. 
The width of these classical
fluctuations
is chosen to be equal to the width of quantum fluctuations in the vacuum
calculated for free fields.
We allow fluctuations in the phase of $\delta\phi$ and fix the 
associated gauge degree of freedom
by choosing $\partial_iA_i=0$.
The longitudinal component of $E_i$ is determined from the Gauss law
(\ref{gauss}).

In the very beginning of our simulation, when the fields are in vacuum,
the
conditions required by the classical approximation are not satisfied,
but we expect that the final results will be unaffected. What is important
is not the precise nature of the initial fluctuations, but that
some small fluctuations are present.

As in the scalar theory \cite{kofman96}, the time evolution
begins with a period of parametric resonance. 
The resonance parameter $q$ is given by $q\approx e^2/\lambda$. Let us first
consider the case $q\gg 1$, in which the resonance is broad, and
during the first
oscillations, a large amount of energy is transferred from
the zero mode $\overline{\phi}$ to the long-wavelength modes
$p\sim \lambda^{1/2}\overline{\phi}_0$ of
$A_i$ and $\delta\phi$, from which it soon
spreads to all modes with
$p\lsim p_* \approx e\overline{\phi}_0$.
We can approximate the state of the system after this period
by assuming that the modes with $p\lsim p_*$ thermalize to
some effective
temperature $T_{\rm eff}$, but those with $p\gsim p_*$ remain in vacuum.
Then the energy density in these fluctuations is 
\begin{equation}
\rho\approx \int^{p_*}\frac{d^3p}{(2\pi)^3}p^2\frac{T_{\rm eff}}{p^2}
\sim p_*^3T_{\rm eff},
\end{equation}
and after preheating it is of the same order as the initial
energy density in the zero mode $\rho_0\sim e^2\overline{\phi}_0^4$,
which implies $T_{\rm eff}\sim \overline{\phi}_0/e$.
In the reheating picture, the temperature after the
equilibration of the fields would be 
$T_r\sim \sqrt{e}\overline{\phi}_0\ll T_{\rm eff}$.

Since the occupation number of the long-wavelength modes is 
$n_p\sim T_{\rm eff}/p$, which is large when $p\lsim p_*$
provided that $e\ll 1$,
the classical approximation works well after preheating
begins.

The zero mode $\overline{\phi}$ continues oscillating around
the minimum, but the fluctuations in
$\delta\phi$ and $A_i$ induce an effective mass term
\begin{equation}
m_{\rm eff}^2\approx
-2\lambda v^{2} + 4\lambda\langle \delta\phi^{2}\rangle
+ e^{2}\langle A_i^{2} \rangle \;.
\label{omameff}
\end{equation} 
The magnitude of the fluctuation terms is
\begin{equation}
\langle \delta\phi^{2}\rangle\sim\langle A_i^{2} \rangle
\sim\int^{p_*}\frac{d^3p}{(2\pi)^3}\frac{T_{\rm eff}}{p^2}
\sim p_*T_{\rm eff}
\sim \overline{\phi}_0^2. 
\end{equation}
In the reheating
picture, the fluctuation terms would be much smaller,
$\langle \delta\phi^{2}\rangle\sim\langle A_i^{2} \rangle\sim T_r^2
\sim e\overline{\phi}_0^2$. This shows that $m_{\rm eff}^2$ can become
positive, thereby restoring the symmetry, even if the reheating temperature
is below $T_c$.

When the Universe expands further, the fluctuation terms decrease and
the system undergoes a phase transition to the broken phase. 
The nature of this transition can be studied by calculating
the effective potential of $\overline{\phi}$ in the background of
the fluctuations $\delta\phi$ and $A_i$.
If $e^2\gg\lambda$,
the contribution from $A_i$ will be more important.
Taking the one-loop contribution from the gauge field
into account, we have
\begin{equation}
V_{\rm eff}(\overline{\phi})\approx
-2\lambda v^2\overline{\phi}^2+\lambda\overline{\phi}^4
+T_{\rm eff}\!\int^{p_*}\!
\frac{d^3p}{(2\pi)^3}\log\frac{p^2+m_A^2}{p^2},
\label{effpot}
\end{equation}
where $m_A\sim e\overline{\phi}$ is the photon mass generated
by the Higgs mechanism.

To understand the shape of the potential (\ref{effpot}), 
we expand it both for small
and large $\overline{\phi}$,
\begin{equation}
V_{\rm eff}(\overline{\phi})
\approx 
\left\{
\begin{array}{lr}
m^2_{\rm eff}\overline{\phi}^2
-C_1e^3T_{\rm eff}\overline{\phi}^3+\lambda\overline{\phi}^4, 
&(\overline{\phi}\ll p_*/e)\\
C_2T_{\rm eff}p_*^3\ln\frac{e\overline{\phi}}{p_*}
-2\lambda v^2\overline{\phi}^2+\lambda\overline{\phi}^4, 
&(\overline{\phi}\gg p_*/e)
\end{array}\right.
\label{effpot2}
\end{equation}
where $C_1$ and $C_2$ are numerical factors and $m^2_{\rm eff}$ is given by
Eq.~(\ref{omameff}).

The origin $\overline{\phi}=0$ is a local minimum whenever 
$m_{\rm eff}^2$ is positive. 
Assuming first that $p_*>ev$, the cubic term in Eq.~(\ref{effpot2})
induces another minimum for the potential when $m_{\rm eff}^2$
becomes small enough, and eventually when this new minimum becomes 
the global one the system enters the Higgs phase in a first
order phase transition. While this phenomenon is present also in 
equilibrium~\cite{Coleman:1973jx}, 
the transition is stronger in our case, since the cubic term is
proportional to $T_{\rm eff}\gg T_r$.
The existence of this minimum requires $e^2\gsim\lambda$, since otherwise
the contribution from the scalar loop, which does not contain
any cubic term, would dominate 
in Eq.~(\ref{effpot2}).

Even if $e^2\lsim\lambda$, the potential can have two minima,
provided that $p_*<ev$~\cite{khlebnikov98}.
Then the tree-level minimum
is the global one if
the logarithmic term in Eq.~(\ref{effpot2})
is smaller than $\lambda v^4$, i.e.~$T_{\rm eff}p_*^3\lsim\lambda v^4$. 
It is difficult to simultaneously satisfy this inequality, along with 
the condition $m^2_{\rm eff} > 0$, and we were unable to do this in our 
simulations.

%the above calculation implies that
%$m_{\rm eff}^2$ is positive and the origin is therefore a minimum of the
%potential. However, if $\overline{\phi}$ is large, the Higgs mechanism 
%generates
%an extra mass term $m_A^2\approx e^2\overline{\phi}^2$ for $A_i$, 
%and if $m_A\gsim p_*$, this
%suppresses $\langle A_i^{2} \rangle$, 
%\begin{equation}
%\langle A_i^{2} \rangle
%\sim\int^{p_*}\frac{d^3p}{(2\pi)^3}\frac{T_{\rm eff}}{p^2+m_A^2}
%\sim \frac{T_{\rm eff}p_*^3}{m_A^2}.
%\end{equation}
%When $T_{\rm eff}$ and $p_*$ decrease, 
%$m_{\rm eff}^2$ eventually becomes negative at some intermediate 
%$\overline{\phi}$,
%and the potential acquires another minimum. When this broken 
%minimum becomes
%the global one, the symmetric phase becomes metastable and eventually
%the system undergoes a first-order transition to the Higgs phase.

\begin{figure}
\epsfig{file=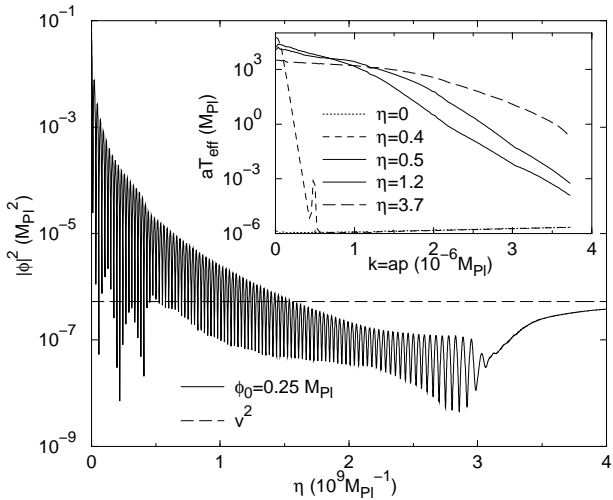,width=8cm}
\caption{The time evolution of $|\phi|^2$ in the first simulation
on a $320^3$ lattice with
the initial condition $\overline{\phi}_0=0.25 M_{\rm Pl}$.
That $|\phi|^2$ is below $v^2$ indicates that the symmetry is restored.
At $\eta\approx 3\cdot 10^9 M_{\rm Pl}^{-1}$, 
the transition to the Higgs phase
takes place.
The inset shows
the effective temperature of different Fourier modes of the
electric field $\tE_i$ measured at various values of $\eta$
(given in units of $10^9 M_{\rm Pl}^{-1}$).
The
energy density in the short-wavelength modes is suppressed
by a factor of $10^4$ relative to the long-wavelength modes
even at the end of the simulation, 
and therefore
the discretization errors are expected to be small.
}
\label{fig:phi2.2}
\end{figure}

To confirm this expected behaviour, we carried out a numerical
lattice simulation with our model.
We chose $\overline\phi_0=0.25 M_{\rm Pl}$ and
$\lambda=2\cdot 10^{-13}$, 
$e=6.4\cdot 10^{-6}$, 
$v=7.2\cdot 10^{-4} M_{\rm Pl}$.
The lattice
spacing was $\delta x=9.3\cdot 10^5 M_{\rm Pl}^{-1}$, %corresponds to 0.15
the time step $\delta\eta=1.2\cdot 10^5 M_{\rm Pl}^{-1}$ %corresponds to 0.02
and
the size of the lattice was $320^3$. The universe was assumed to be radiation
dominated with $a(\eta)=1+\eta H$, where 
$H=8.3\cdot 10^{-8}M_{\rm Pl}$. %0.51
Since $\overline{\phi}$ is not
a gauge-invariant quantity and can therefore only be defined in the vacuum,
we did not measure its value. Instead, we show $|\phi|^2$ as a function
of time in Fig.~\ref{fig:phi2.2}. 
The fact that $|\phi|^2<v^2$ when
$1.5\cdot10^{9}M_{\rm Pl}^{-1}\lsim\eta\lsim 3\cdot 10^{9}M_{\rm Pl}^{-1}$
clearly shows that the gauge symmetry is restored and the system
is in the Coulomb phase.
The amplitude of the oscillations remains quite large, which is
probably a finite-size effect. In an infinite system, there would be more
infrared modes to which the zero mode of $\phi$ could decay.
Eventually, the system
enters the Higgs phase in a first-order transition, as in
the scalar theory. The first-order nature of the transition can be seen
from the configurations during the transition; for example by looking at the
isosurface of $|\phi|^2$ 
we would see a growing bubble of the Higgs phase characterized by a larger
value of $|\phi|^2$.

\begin{figure}
\epsfig{file=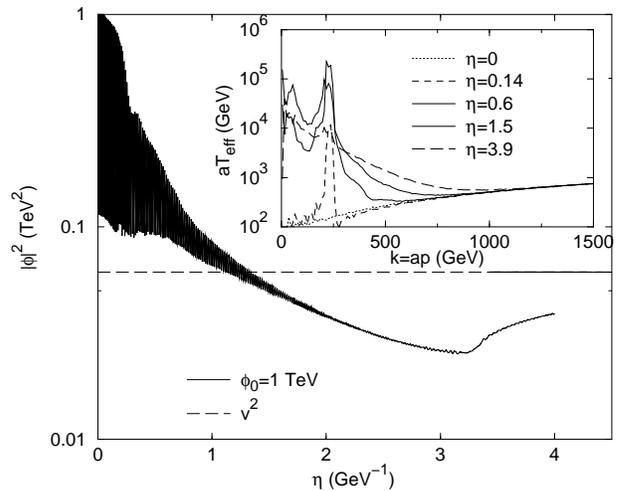,width=8cm}
\caption{The time evolution of $|\phi|^2$ in the second simulation
on a $240^3$ lattice with
the initial condition $\overline{\phi}_0=1$~TeV.
Again, the symmetry is restored, and
at $\eta\approx 3$~GeV$^{-1}$, 
the transition to the Higgs phase
takes place.
The effective temperature measured at various $\eta$ (given in units of
GeV$^{-1}$) in the inset shows that
in this case the resonance is narrower, but the temperature still grows
to high values.
}
\label{fig:phiew}
\end{figure}

In order to check that the separation of scales below and above $p_*$
indeed takes place, we measured the effective temperature of
different Fourier modes of the 
electric fields $E_i$
at various times during the simulation. 
A reason for choosing this quantity
rather than the power spectrum of
$\tphi$ or $A_i$ is that it is gauge-invariant. In equilibrium
$|\tE_{i,k}|^2$ is constant and its magnitude is proportional to
the temperature. Therefore we can use it to define
the effective temperature of a single mode 
\begin{equation}
T_{\rm eff}(p=k/a)=\frac{1}{2a}|\tE^{\rm T}_{i,k}|^2\frac{d^3k}{(2\pi)^3},
\end{equation}
where the superscript T indicates that we have included only the transverse
component $k_i\tE^{\rm T}_{i,k}=0$, since the longitudinal component is fixed
by the Gauss law.
The inset of
Fig.~\ref{fig:phi2.2} shows the product $aT_{\rm eff}$, which
is the effective temperature of the rescaled fields, as a
function of the conformal momentum $k=pa$.

Immediately after preheating,
the temperature of the long-wavelength modes is 
$T_{\rm eff}\sim 10^4M_{\rm Pl}$ and
the occupation number $n_p=T_{\rm eff}(p)/p\sim10^{10}$
is huge.
The cutoff momentum is $p_*=k_*/a\approx 10^{-6} M_{\rm Pl}/a$. 
With time, the modes with higher and higher $k$ thermalize and the
temperature decreases, but since the couplings are small, this process
is very slow.
The modes with
$k\gg k_*$ are strongly suppressed even after the phase transition, and
therefore we believe that the lattice approximation remains reliable even
at the end of the simulation.
Because the modes with the
highest momenta do not remain exactly in the vacuum, 
discretization errors cannot be ruled out completely.

For the electroweak theory, the opposite case $q<1$ is more relevant, since
$q\approx m_W^2/m_H^2$. In this case, the parametric resonance is narrow
and the energy transfer is less efficient. However, since the expansion rate
of the Universe is much slower in this case, it could still lead to a
similar phenomenon. Most of the energy of the inflaton
is transferred to a narrow
momentum range of the fluctuations, but the long-wavelength modes thermalize
and the energy is spread to all long-wavelength modes. After that, the
system should behave as in the case with a broad resonance. In equilibrium,
the phase transition is not of first order if $e^2>\lambda$, but as
discussed earlier, we expect the
transition to be stronger in our case. 

The realistic values for the couplings in the electroweak theory would be
$\lambda\sim e^2\sim 1$, but in that case our simulations are not reliable.
With these couplings, the interactions are important even in the vacuum
state, and the classical approximation cannot be trusted.
Therefore, we have used slightly smaller couplings,
$\lambda=0.04$ and $e=0.14$, 
which allow us to use the classical approximation.
The initial value of the Higgs field was $\overline{\phi}_0=1$~TeV.
We also chose $v=246$~GeV and $a(\eta)=1+\eta H$ with $H=0.7$~GeV.
The lattice spacing was $\delta x=1.4$~TeV$^{-1}$, time step 
$\delta\eta=0.14$~TeV$^{-1}$, and the lattice size $240^3$.

In this case, $\phi$ cannot be the inflaton, because its couplings are
much too strong. However, 
the homogeneous initial condition for $\phi$ may arise
from a previous preheating phase, in which $\phi$ couples to the inflaton
with a coupling constant that is much smaller than $e$. Then
the parametric resonance will transfer a large amount of energy
to modes of $\phi$ with very long wavelengths. The alternative possibility
is that quantum fluctuations of $\phi$ give it a large spatial average
during inflation.

As in the earlier case
for the inflaton, we show $|\phi|^2$ and the effective temperature
of different modes of $E_i$ in Fig.~(\ref{fig:phiew}). This time, the 
energy is
transferred into a narrow band of gauge field modes. Nevertheless,
the long-wavelength modes thermalize, and we reach a similar situation
to that
in the first case, in which the long-wavelength modes 
$k\lsim 300$~GeV have an
effective temperature $T_{\rm eff}\approx 10^4$~GeV, and the symmetry is
restored. 
At $\eta\approx 3$~GeV$^{-1}$, the system undergoes a phase transition
to the Higgs phase. The transition is
not of first order, but it is still rather sharp.

In the electroweak theory, the conservation of
baryon number would be violated by sphaleron
configurations with a rate 
$\Gamma_{\rm sph}\sim \alpha_W^5T_{\rm eff}^5/p_*$~\cite{arnold}
when the symmetry is temporarily restored, 
and as discussed in Ref.~\cite{Garcia-Bellido:2000px},
the oscillations of the Higgs field could create a large baryon asymmetry. 
If the transition to the Higgs phase is sharp enough, the baryon number 
violation ceases instantaneously, and the produced baryon asymmetry
remains.

Our simulations show that a gauge-Higgs system exhibits 
the same behaviour as the scalar model considered in 
Ref.~\cite{khlebnikov98}. The first case we considered shows
that a non-thermal phase transition is possible
if the inflaton is charged under a gauge group.
Although we restricted ourselves to an Abelian model, we believe that the 
qualitative features of our results would be the
same in non-Abelian theories. In many models, this phase transition would
lead to the formation of cosmic strings or other topological defects.

The second case we considered has the qualitative features of the
electroweak theory, and we find that the symmetry gets restored
although the parametric resonance is narrow, provided that the expansion
of the Universe is slow enough. Unlike in the standard
thermal phase transition scenario, the transition to
the Higgs phase is sharp, which makes it possible to preserve the
produced baryon asymmetry. This supports the
picture of electroweak baryogenesis at preheating.

%%%%%%%%%%%%%%%%%%%%%%%%%%%%%%%%%%%%%%%%%%%%%%%%%%%%%%%%
\acknowledgements
We would like to thank Andrei Linde and Mark Hindmarsh
for useful conversations. 
EJC and AR are supported by PPARC, and AR also partly by the University
of Helsinki. This work was conducted on the SGI Origin platform using COSMOS
Consortium facilities, funded by HEFCE, PPARC and SGI.

\end{document}